\documentclass[aps, prd, onecolumn, tightenlines, notitlepage, superscriptaddress, nofootinbib, preprintnumbers, floatfix,showkeys]{revtex4}

\usepackage[normalem]{ulem}
\usepackage{amstext}
\usepackage{amssymb}
\usepackage{amsmath}
\usepackage{graphicx}
\usepackage{url}
\usepackage{color}
\usepackage{ulem}
\usepackage[utf8]{inputenc}
\pdfoutput=1
\usepackage{textcomp}

\usepackage{epsfig,amsfonts,mathrsfs,amsmath,amssymb,graphicx,color,slashed,multirow}
\usepackage{amsmath,latexsym,amssymb,graphicx,slashed,color,enumerate,url,cancel,gensymb}
\usepackage{textcomp}

\usepackage[x11names]{xcolor}
\usepackage[colorlinks]{hyperref}

\hypersetup{colorlinks,citecolor= nicered,linkcolor= nicered}
\definecolor{nicered}{rgb}{0.7,0.1,0.1}
\definecolor{nicegreen}{rgb}{0.1,0.5,0.1}

\AtBeginDocument{\hypersetup{citecolor=nicered,linkcolor=blue,urlcolor=nicegreen}}

\begin{document}

\title{{\Large Probing neutrino quantum decoherence at reactor experiments}}

\author{Andr\'e de Gouv\^ea}\email{degouvea@northwestern.edu}
\affiliation{Northwestern University, Department of Physics \& Astronomy, 2145 Sheridan Road, Evanston, IL 60208, USA}
\author{Valentina De Romeri}\email{deromeri@ific.uv.es}
\affiliation{ Institut de F\'{i}sica Corpuscular CSIC/Universitat de Val\`{e}ncia, Parc Cient\'ific de Paterna\\
 C/ Catedr\'atico Jos\'e Beltr\'an, 2 E-46980 Paterna (Valencia) - Spain}
\author{Christoph A. Ternes}\email{chternes@ific.uv.es}
\affiliation{ Institut de F\'{i}sica Corpuscular CSIC/Universitat de Val\`{e}ncia, Parc Cient\'ific de Paterna\\
 C/ Catedr\'atico Jos\'e Beltr\'an, 2 E-46980 Paterna (Valencia) - Spain}

\begin{abstract}

We explore how well reactor antineutrino experiments can constrain or measure the loss of quantum coherence in neutrino oscillations. We assume that decoherence effects are encoded in the size of the neutrino wave-packet, $\sigma$. We find that the current experiments Daya Bay and the Reactor Experiment for Neutrino Oscillation (RENO) already constrain $\sigma>1.0\times 10^{-4}$~nm and estimate that future data from the Jiangmen Underground Neutrino Observatory (JUNO) would be sensitive to  $\sigma<2.1\times 10^{-3}$~nm.
If  the effects of loss of coherence are within the sensitivity of JUNO, we expect $\sigma$ to be measured with  good precision. 
The discovery of nontrivial decoherence effects in JUNO would indicate that our understanding of the coherence of neutrino sources is, at least, incomplete.
\end{abstract}

\preprint{NUHEP-TH/20-01}
\maketitle

\section{Introduction}

Neutrino oscillations are a consequence of the fact that neutrinos are produced as coherent quantum superpositions of the different neutrino mass eigenstates ($\nu_1,\nu_2,\nu_3$, with masses $m_1,m_2,m_3$, respectively). 
This, in turn, is a consequence of the fact that the charged-current weak interactions are not diagonal in the basis of the mass eigenstates for both the charged leptons and the neutrinos. 
In other words, in the basis where the charged-lepton masses are diagonal, the neutrino interaction eigenstates ($\nu_e,\nu_{\mu},\nu_{\tau}$) are linear superpositions of the mass eigenstates: $\nu_{\alpha}=U_{\alpha i}\nu_i$, $\alpha=e,\mu,\tau$, $i=1,2,3$ and $U_{\alpha i}$, are the elements of the unitary leptonic-mixing matrix. 

That many neutrino sources are coherent is not a trivial statement. 
It is, ultimately, a consequence of the fact that, compared to the typical energy and distance scales involved in neutrino production and detection, neutrino masses are all tiny and neutrino wave-packets are large. 
The coherence of neutrino sources is the subject of a lot of confusion in and outside the neutrino physics community but has also been discussed very proficiently in the literature, for example~\cite{Nussinov:1976uw,Kayser:1981ye,Giunti:1991ca,Giunti:1991sx,Kiers:1997pe,Giunti:1997wq,Grimus:1998uh,Akhmedov:2010ms,Naumov:2010um,Akhmedov:2012uu}. 
Here, we will not add to this fascinating issue. 

Neutrinos from the Sun detected on the surface of the Earth are best described, for various reasons, as incoherent superpositions of mass eigenstates. 
The same is expected of, for example, neutrinos produced in supernova explosions and detected on the surface of the Earth. 
On the other hand, many neutrino sources, including all terrestrial sources and neutrinos produced in the atmosphere, are treated as perfectly coherent. 
To date, this has proven to be an excellent approximation, in agreement with our best detailed understanding of neutrino production and corroborated by the oscillations interpretation of data from neutrino experiments. 

Nonetheless, neutrino sources cannot be indiscriminately coherent. 
At least in principle, one can imagine circumstances that lead to neutrino sources that are ``partially coherent'' and all neutrino ``beams'' are expected to lose coherence as a function of the neutrino proper time. 
To date, however, there is no experimental evidence of distance-dependent loss of coherence for propagating neutrinos. 
This is the subject of this manuscript. 

The loss of coherence does not prevent neutrino flavor-change but, instead, ``smooths out'' the oscillatory behavior of the neutrino oscillation phenomenon as the neutrinos move away from the source. 
Here, we explore whether high-resolution, high-statistics measurements of the flux of antineutrinos produced in nuclear reactors are sensitive to neutrino decoherence or can be used to place meaningful bounds on how coherent nuclear reactors are as neutrino sources. 
Nuclear reactors are excellent laboratories to study neutrino coherence. 
They are a compact source of electron antineutrinos (few meters compared to the neutrino oscillation lengths, which are of order kilometers to hundreds of kilometers), and the neutrino energies can be measured with great precision in relatively compact detectors (several meters in size but centimetric position resolution). 

We are particularly interested in data from the Jiangmen Underground Neutrino Observatory (JUNO), currently under construction. 
The JUNO baseline is chosen in order to maximize sensitivity to the neutrino mass-ordering and is much longer than the oscillation length due to the ``atmospheric'' mass squared difference. %
Nonetheless, the energy and position resolutions are such that ``atmospheric'' oscillations are visible, rendering JUNO uniquely well suited to probe decoherence effects.  
 
In Sec.~\ref{sec:osc}, we introduce neutrino oscillations and discuss the formalism we will use to describe and constrain decoherence, concentrating on how it modifies the neutrino oscillation probabilities at reactor experiments. 
In Sec.~\ref{sec:current}, we analyze data from the ongoing Reactor Experiment for Neutrino Oscillation (RENO) and the Daya Bay reactor neutrino experiment, and discuss bounds on the wave-packet width, introduced in Sec.~\ref{sec:osc}. 
In Sec.~\ref{sec:JUNO}, we discuss the sensitivity of JUNO. 
We summarize our results in Sec.~\ref{sec:conc}, and offer some concluding remarks. 

\section{Neutrino oscillations, including decoherence}
\label{sec:osc}

Nuclear reactors produce an intense flux of electron antineutrinos with energies roughly in the $[1 - 8]$~MeV range. These are detected some distance away from the source via inverse beta-decay, which allows one to measure the neutrino energy on an event-by-event basis with good precision. If the flux of electron antineutrinos is, somehow, known, reactor neutrino oscillation experiments can measure the survival probability of electron antineutrinos, $P(\bar{\nu}_e \to \bar {\nu}_e)$, as a function of energy and baseline.

It is straight forward to compute $P(\bar{\nu}_e \to \bar {\nu}_e)$. Here we include, rather generally, the effects of decoherence among the mass eigenstates. For a fixed neutrino energy $E$ and baseline $L$, the density matrix $\rho_{jk}$, $j,k=1,2,3$, of the antineutrino state produced in the nuclear reactor, in the mass basis, can be written as 
\begin{equation}
\rho_{jk}(L,E)  = U_{ej}^*U_{ek}\exp[-i\Delta_{jk}]\exp[-\xi_{jk}(L,E)]\,,
\end{equation}
where 
\begin{equation}
\Delta_{jk}\equiv 2\pi\frac{L}{L^{\rm osc}_{jk}} \equiv \frac{\Delta m^2_{jk}L}{2E}\,,
\end{equation} 
$\Delta m^2_{jk}=m_j^2-m^2_k$, and $\xi_{jk}(L,E)=\xi_{kj}(L,E)$ quantifies the loss of coherence as a function of the neutrino energy and the baseline. In the absence of decoherence, $\xi_{jk}=0$. The survival probability, including decoherence effects, is simply the $ee$ element of the density matrix and reads
\begin{align}
  P^\text{dec}(\overline{\nu}_e \to \overline{\nu}_e) =
    \sum_{j,k} |U_{e j}|^2 |U_{e k}|^2 \,
        \exp[
          - i \Delta_{jk}
          - \xi_{jk}] \,,
\end{align}
or
\begin{align}
  1- P^\text{dec}(\overline{\nu}_e \to \overline{\nu}_e) &= 2|U_{e1}|^2|U_{e2}|^2\left(1- \cos\left(\frac{\Delta m_{21}^2 L}{2 E}\right) e^{-\xi_{21}}\right) \nonumber
\\& +2|U_{e1}|^2|U_{e3}|^2 \left(1-  \cos\left(\frac{\Delta m_{31}^2 L}{2 E}\right) e^{-\xi_{31}} \right) \nonumber
\\& +2|U_{e2}|^2|U_{e3}|^2 \left(1-  \cos\left(\frac{\Delta m_{32}^2 L}{2 E}\right)  e^{-\xi_{32}}\right)\,. 
\label{eq:Pee_dec}    
\end{align}
It is trivial to see that we recover the standard expression for the electron antineutrino disappearance when all $\xi_{jk}\to 0$. Throughout, we will use the standard PDG parameterization of the leptonic mixing matrix where $|U_{e1}|^2=\cos^2\theta_{12}\cos^2\theta_{13}$, $|U_{e2}|^2=\sin^2\theta_{12}\cos^2\theta_{13}$, and $|U_{e3}|^2=\sin^2\theta_{13}$, and, unless otherwise noted, we assume that the true values of the relevant neutrino oscillation parameters are
\begin{eqnarray}
& \Delta m_{31}^2 = 2.5\times10^{-3}~\text{eV}^2, & \Delta m_{21}^2 = 7.55\times10^{-5}~\text{eV}^2, \nonumber \\ 
& \sin^2\theta_{13} = 0.0216, &  \sin^2\theta_{12} = 0.32, \label{eq:params}
\end{eqnarray}
in agreement with the best-fit values obtained from the world's neutrino data~\cite{deSalas:2017kay}. We assume the neutrino mass-ordering is normal ($\Delta m^2_{31}>0$) and assume this information is known. We will comment on the consequences of this assumption when relevant.

Different physical effects lead to decoherence~\cite{Kiers:1995zj,Ohlsson:2000mj,Beuthe:2001rc,Beuthe:2002ej,Giunti:2003ax,Blennow:2005yk,Farzan:2008eg,Kayser:2010pr,Naumov:2013uia,Jones:2014sfa,Akhmedov:2019iyt,Grimus:2019hlq,Naumov:2020yyv}. Here, we will concentrate on decoherence effects that grow as the baseline grows and parameterize the decoherence parameters as~\cite{Giunti:1991sx,Beuthe:2002ej,Kayser:2010pr}
\begin{equation}
\xi_{jk}(L,E)=\bigg( \frac{L}{L^{\rm coh}_{jk}} \bigg)^2,
\label{eq:dec_fac}
\end{equation}
and further parameterize the coherence lengths as~\cite{Giunti:1991sx,Beuthe:2002ej,Kayser:2010pr}
\begin{equation}
  L^{\rm coh}_{jk} = \frac{4 \sqrt{2} E^2}{|\Delta m_{jk}^2|} \sigma\,. 
  \label{eq:Lcoh}
\end{equation}
Concretely, as discussed in \cite{Giunti:1991sx,Beuthe:2002ej,Kayser:2010pr}, $\sigma$ is the width of the neutrino wave-packet and depends on the properties of the neutrino source  and of the detector. The physics that leads to this type of decoherence is the fact that the different neutrino mass eigenstates propagate with different speeds and, given enough time, the wave-packets ultimately separate. From a more pragmatic point of view, here $\sigma$ is the single parameter that characterizes the effects of decoherence, and has dimensions of length. Decoherence effects vanish as the coherence lengths become very long: $\sigma\to\infty$, and we highlight that the different coherence lengths are inversely proportional to the associated neutrino mass-squared differences. Here, constraining neutrino decoherence assuming the data are consistent with a perfectly coherent beam is equivalent to placing a lower bound on $\sigma$.  There are other sources of decoherence, including those associated to the production or detection of the neutrinos. The latter, for example, are independent of the baseline $L$ and will not be considered here. Recent constraints are discussed in \cite{{An:2016pvi}}.

\begin{figure}[t]
\centering
\includegraphics[width=0.55\textwidth]{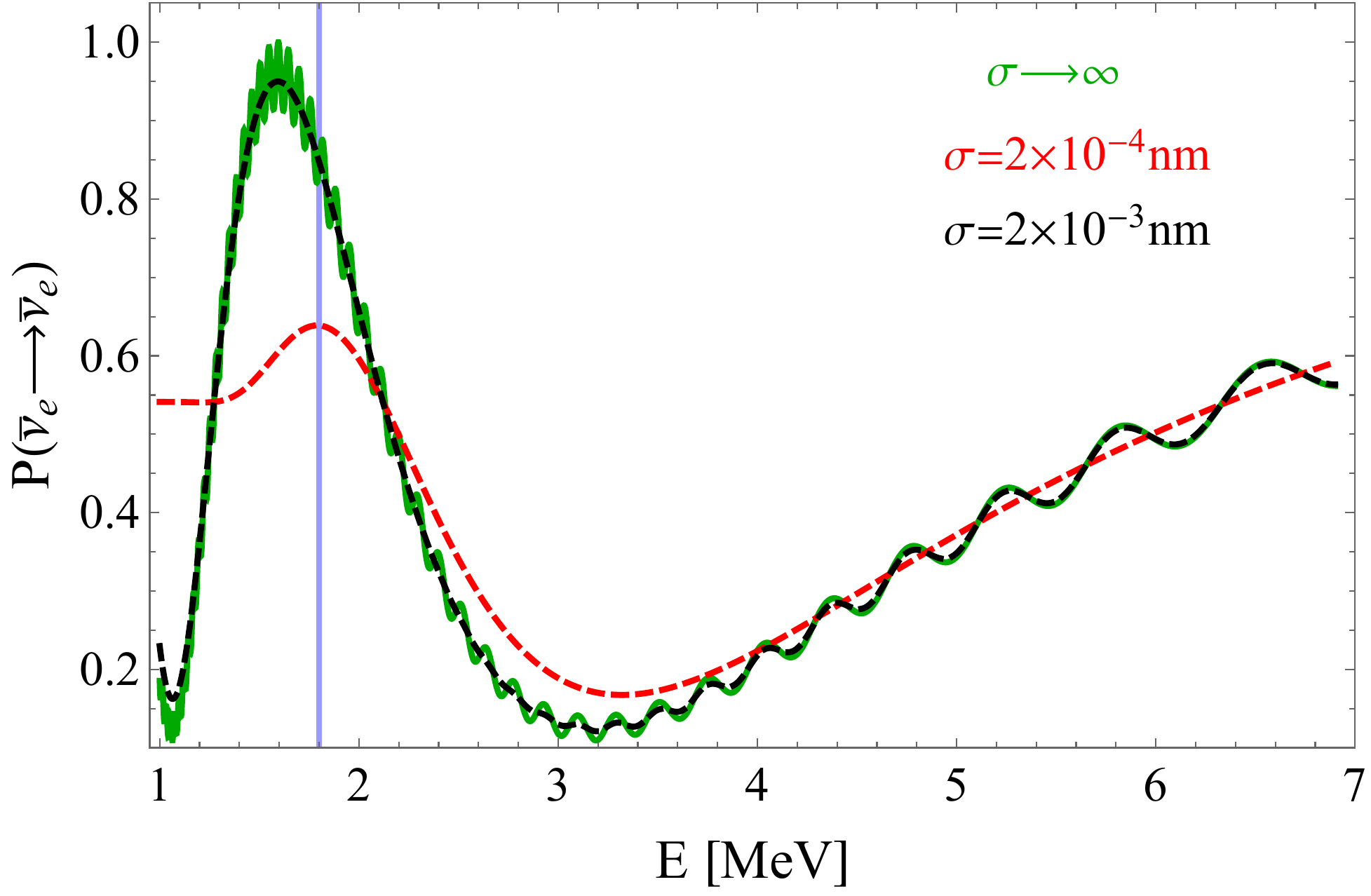}
\caption{The electron antineutrino oscillation probability as a function of the neutrino energy for the JUNO average baseline of $L=52.5$~km for different values of the decoherence parameter $\sigma$. The values of the standard oscillation parameters are listed in Eq.~(\ref{eq:params}). The blue, vertical line indicates the threshold for inverse beta-decay.}
\label{fig:osc_prob}
\end{figure}

Decoherence effects in reactor experiments grow with the baseline and decrease with the neutrino energy. Fig.~\ref{fig:osc_prob} depicts the expected $\bar {\nu}_e \to \bar {\nu}_e$ oscillation probability for typical reactor neutrino energies and the JUNO average baseline $L=52.5$~km, assuming the oscillation parameters are the ones in Eqs.~(\ref{eq:params}). The green, solid curve corresponds to standard oscillations with no decoherence effects while the red and black dashed ones are the expected disappearance probabilities in the presence of decoherence effects with $\sigma = 2 \times 10^{-4}$~nm and $\sigma = 2 \times 10^{-3}$~nm, respectively. Note that the fast oscillations ``disappear'' first and that the effect is more pronounced at smaller neutrino energies. 

\section{Current Constraints from RENO and Daya Bay}
\label{sec:current}

RENO and Daya Bay are reactor neutrino experiments in South Korea and China, respectively, that measure the flux of antineutrinos from nuclear reactors at $L\sim 100$~m and $L\sim 1$~km, using information from both the near and far detectors to measure $P(\overline{\nu}_e \to \overline{\nu}_e)$. 
Given the typical reactor neutrino energies and the 1~km baselines, these experiments are sensitive to $\Delta m^2_{31}$ and $\sin^2\theta_{13}$ but insensitive to the ``solar'' parameters  $\Delta m^2_{21}$ and $\sin^2\theta_{12}$. 
This effective-two-flavor approximation also applies to the decoherence effect since $L_{12}^{\rm coh}\gg L_{13}^{\rm coh}\simeq L_{23}^{\rm coh}$. 
Hence, at the relevant energies and baselines, 
\begin{equation}
  1- P^\text{dec}(\overline{\nu}_e \to \overline{\nu}_e) = \frac{1}{2}\sin^22\theta_{13}  \left[1-  \cos\left(\frac{\Delta m_{31}^2 L}{2 E}\right) \exp\left(-\bigg( \frac{L}{L^{\rm coh}_{13}} \bigg)^2\right) \right]\,,
\label{eq:Pee_dec_2flavor}    
\end{equation}
is an excellent description of electron antineutrino disappearance at RENO and Daya Bay.\footnote{In our numerical calculations, we use the full three-neutrino description, Eqs.~\eqref{eq:Pee_dec}, \eqref{eq:dec_fac} and \eqref{eq:Lcoh}.} 
For the same reasons, data from RENO and Daya Bay are insensitive to the neutrino mass-ordering and the results presented here do not depend on our assumption that the neutrino mass-ordering is normal.

RENO uses a power plant with six nuclear reactors as neutrino sources and consists of two identical detectors at two different locations. Daya Bay makes use of six nuclear reactors located at two nearby sites. In the case of Daya Bay, there are eight identical detectors located at three different experimental halls; two experimental halls contain two detectors each that serve as near detectors, while the remaining four detectors are in the third experimental hall, which is further away.

For the results presented here, we use the most up-to-date data from the two experiments, corresponding to 2900 days of data from RENO~\cite{RENO-neutrino2020} and 1958 days of data from Daya Bay~\cite{Adey:2018zwh}. The necessary information on all technical details, including the baselines, thermal power, fission fractions, and efficiencies, is obtained from Refs.~\cite{Ahn:2010vy,Seo:2016uom,Bak:2018ydk,RENO-neutrino2020} for RENO and Refs.~\cite{An:2016srz,An:2016ses,Adey:2018zwh} for Daya Bay. 
In our statistical analyses, we account for several sources of systematic uncertainties.
We include uncertainties related to the thermal power for each core and to the detection efficiencies, uncertainties on the fission fractions, a shape uncertainty for each energy bin in our analyses, and an uncertainty on the energy scale.

We define the $\chi^2$ function for RENO as
\begin{equation}
 \chi^2_\text{RENO}(\vec{p}) = \min_{\vec{\alpha}}\left\{\sum_{i=1}^{N_\text{RENO}}\left(\frac{R^{F/N}_{\text{dat},i} - R^{F/N}_{\text{exp},i}(\vec{p},\vec{\alpha})}{\sigma^\text{RENO}_i}\right)^2 + \sum_k \left(\frac{\alpha_k - \mu_k}{\sigma_k} \right)^2\right\}\,.
\end{equation}
Here, $R^{F/N}_i = F_i/N_i$, where $F_i$ and $N_i$ are the event numbers in the $i$th energy bin at the far and near detector, respectively. 
$R_{\text{dat},i}$ are the background-subtracted observed event ratios, while $R_{\text{exp},i}(\vec{p},\vec{\alpha})$ are the expected event ratios for a given set of oscillation parameters $\vec{p}$. 
The uncertainty for each bin is given by $\sigma^\text{RENO}_i$.
The last term contains penalty factors for all of the systematic uncertainties $\alpha_k$ with expectation value $\mu_k$ and standard deviation $\sigma_k$. 
Finally the number of bins is given by $N_\text{RENO}$. 

Similarly, for Daya Bay, we define

\begin{equation}
 \chi^2_\text{DB}(\vec{p}) = \min_{\vec{\alpha}}\left\{\sum_{i=1}^{N_\text{DB}}\left(\frac{R^{F/N_1}_{\text{dat},i} - R^{F/N_1}_{\text{exp},i}(\vec{p},\vec{\alpha})}{ \sigma_i^{F/N_1}}\right)^2  + \sum_{i=1}^{N_\text{DB}}\left(\frac{R^{N_2/N_1}_{\text{dat},i} - R^{N_2/N_1}_{\text{exp},i}(\vec{p},\vec{\alpha})}{\sigma_i^{N_2/N_1}}\right)^2 +  \sum_k \left(\frac{\alpha_k - \mu_k}{\sigma_k} \right)^2\right\}\,.
\end{equation}
Here, we take the ratios between the far and the first near detector and between the two near detectors, as was done in Ref.~\cite{Esteban:2018azc}. 
To calculate the expected number of events and the $\chi^2$ functions for each experiment, we use GLoBES~\cite{Huber:2004ka,Huber:2007ji}. 
We use reactor fluxes as parameterized in Ref.~\cite{Huber:2004xh} and the inverse beta-decay cross section from Ref.~\cite{Vogel:1999zy}. 
We analyze the data from each experiment independently and also perform combined analyses, where we use 
\begin{equation}
 \chi^2_\text{COMB}(\vec{p}) = \chi^2_\text{RENO}(\vec{p}) + \chi^2_\text{DB}(\vec{p})\,.
\end{equation}

In order to validate our treatment of the two data sets, we first assume a perfectly coherent source and compare our results to those published by RENO and Daya Bay. Hence, we first consider the case $\vec{p} = (\Delta m_{31}^2, \theta_{13})$.  The solar parameters are fixed to $\sin^2\theta_{12} = 0.32$ and $\Delta m_{21}^2 = 7.55\times 10^{-5}~\text{eV}^2$~\cite{deSalas:2017kay}. As already mentioned, this choice is inconsequential for the results presented here.
\begin{figure}[t]
\centering
\includegraphics[width=1.0\textwidth]{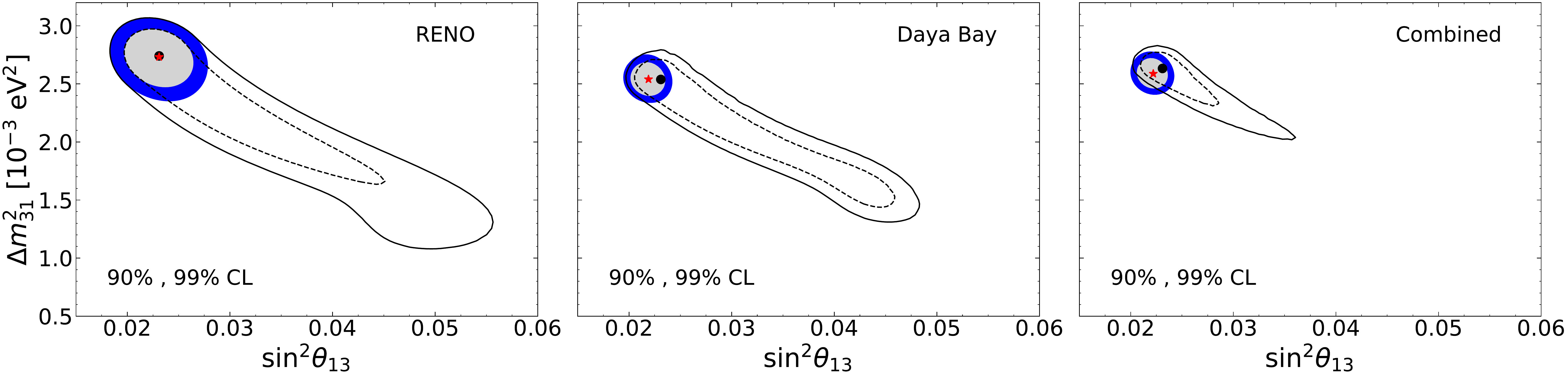}
\caption{
90 and 99\% CL (2 d.o.f.) allowed regions in the $\sin^2\theta_{13}$--$\Delta m^2_{31}$ plane for RENO (left), Daya Bay (center) and the combination of both experiments (right). Filled regions correspond to the analyses assuming a perfectly coherent source, while black lines are obtained after marginalizing over $\sigma$. The best-fit points from the standard analyses are indicated with a red star, while the best-fit values from the analyses including $\sigma$ are denoted by black dots.}
\label{fig:sq13_dm31}
\end{figure}
The grey and blue ellipses in Fig.~\ref{fig:sq13_dm31} correspond to the region of the oscillation parameter space consistent, at the 90 and 99\% CL respectively (for two degrees of freedom), with data from RENO (left) and Daya Bay (center), along with the combined result (right). The combined analysis is clearly dominated by the Daya Bay data. These results agree quantitatively very well with those presented in Refs.~\cite{Bak:2018ydk} and \cite{Adey:2018zwh}.

Next, we allow for the possibility that the wave-packet width $\sigma$ is not infinite, and extend the set of model parameters: $\vec{p} = (\Delta m_{31}^2, \theta_{13}, \sigma)$. 
Marginalizing over $\sigma$, the regions of the $\sin^2\theta_{13}$--$\Delta m_{31}^2$ parameter space consistent with the different data sets are depicted in Fig.~\ref{fig:sq13_dm31} as closed, empty contours (solid at the 99\% CL, dashed at the 90\% CL).
Not surprisingly, the allowed regions on the $\Delta m_{31}^2$--$\sin^2\theta_{13}$ plane are larger once one allows for finite $\sigma$ values. The region of parameter space in the ``combined'' case is noticeably smaller than that allowed by Daya Bay data. 
This is a consequence of the fact that the $L/E$ values probed by RENO and Daya Bay are slightly different and the shapes of the allowed regions are slightly different. 
In particular, the best-fit point in the case of Daya Bay shifts more than that of RENO once finite values of $\sigma$ are allowed. 
The result depicted in Fig.~\ref{fig:sq13_dm31} (center) is in qualitative agreement with the results obtained by the Daya Bay collaboration in Ref.~\cite{An:2016pvi} using a smaller data set~\cite{An:2015rpe}. 
(Note that our definition of $\sigma$ (in coordinate space) is consistent with that of $\sigma_x$ in Ref.~\cite{An:2016pvi}.) 

\begin{figure}[t]
\centering
\includegraphics[width=0.49\textwidth]{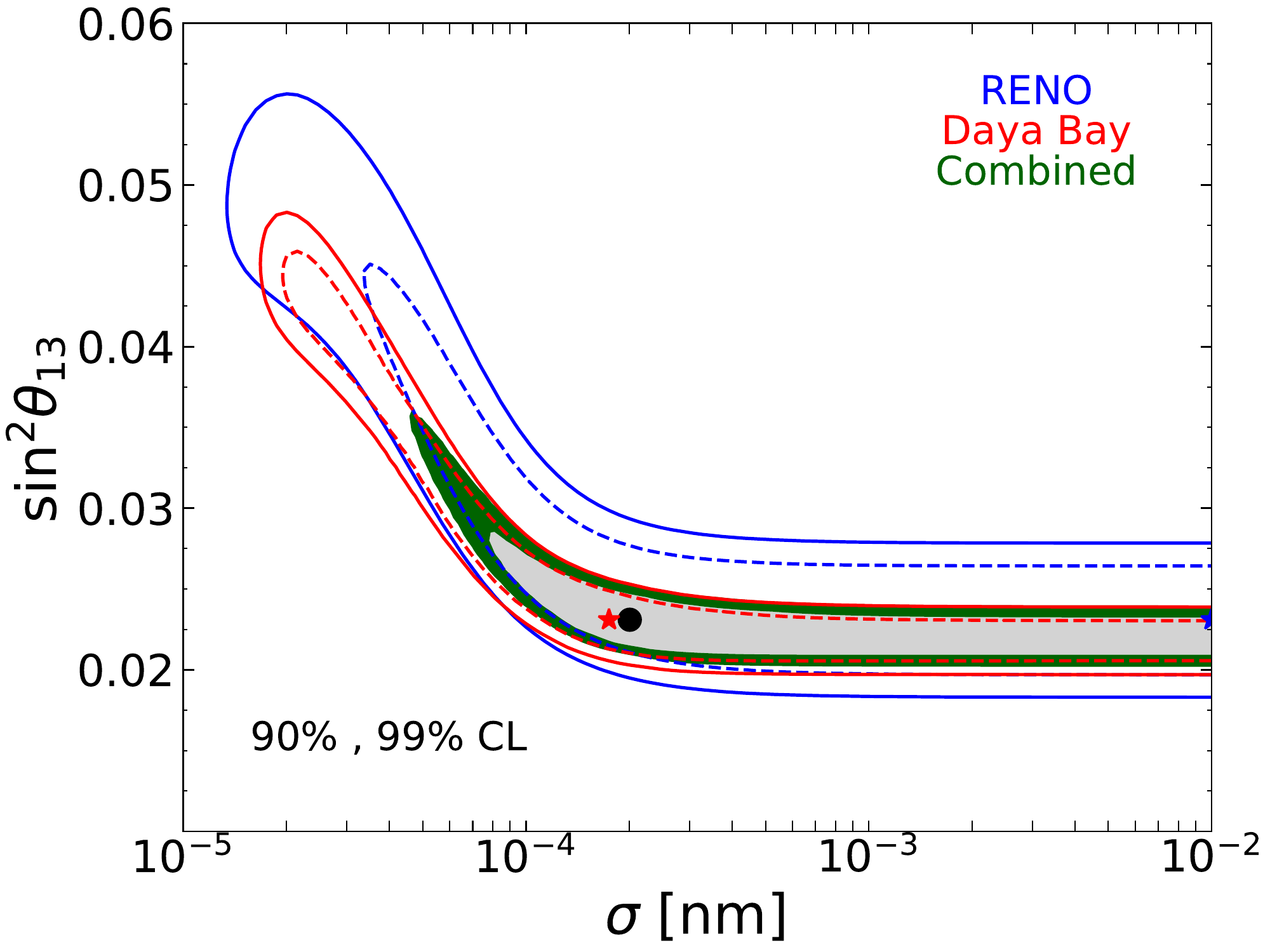}
\includegraphics[width=0.49\textwidth]{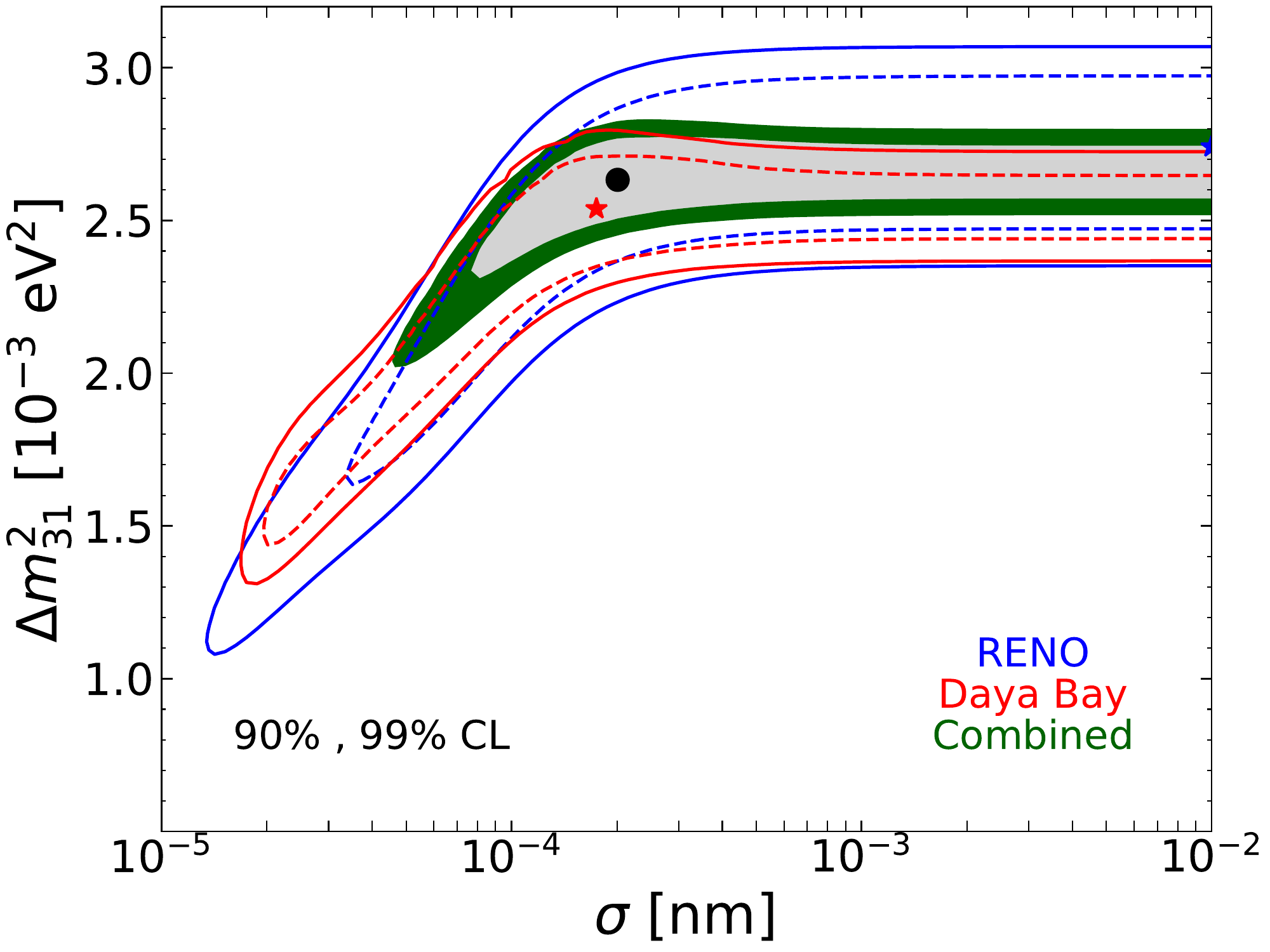}
\caption{
90 and 99\% CL (2 d.o.f.) allowed regions in the $\sigma$--$\sin^2\theta_{13}$ (left) and $\sigma$--$\Delta m^2_{31}$ (right) planes  for RENO (blue lines), Daya Bay (red lines) and the combination of the two (filled regions). Stars denote the best-fit values from the analysis of a single experiment on its own (the best-fit value for RENO lies at $\sigma \sim 10^{-2}$~nm), while the black dot is the best-fit point obtained from the combined analysis.}
\label{fig:sigma_X}
\end{figure}

Fig.~\ref{fig:sigma_X} (left) and Fig.~\ref{fig:sigma_X} (right) depict the allowed regions of the $\sigma$--$\sin^2\theta_{13}$ and $\sigma$--$\Delta m_{31}^2$ parameter spaces, respectively, marginalizing over the absent parameter.
For small enough values of $\sigma$, there is a clear (anti)correlation between $\sigma$ and  $\Delta m_{31}^2$ ($\sin^2\theta_{13}$). 
These correlations are also manifest in the anticorrelation between $\Delta m_{31}^2$ and $\sin^2\theta_{13}$ observed in Fig.~\ref{fig:sq13_dm31}. 

It is straight forward to understand qualitatively the allowed regions in Fig.~\ref{fig:sigma_X} (left) and Fig.~\ref{fig:sigma_X} (right). Both RENO and Daya Bay probe $L/E$ values that include the first oscillation maximum associated to $\Delta  m^2_{31}$ while all other maxima are outside the reach of the two experiments. 
Decoherence effects ``flatten'' the oscillation maximum, an effect that can be partially compensated by increasing $\sin^2\theta_{13}$. 
Hence, for smaller values of $\sigma$ (stronger decoherence), one can obtain a decent fit to the data by increasing $\sin^2\theta_{13}$ relative to the value obtained in the perfectly-coherent hypothesis. 
Decoherence effects also shift the position of the first oscillation maximum to smaller $L/E$ values. 
This is simple to understand and is well illustrated in the red, dashed curve in Fig.~\ref{fig:osc_prob}. 
This can be compensated by lowering the size of $\Delta m^2_{31}$ (longer wave-length). 
Hence, for smaller values of $\sigma$ (stronger decoherence), one can obtain a decent fit to the data by decreasing $\Delta m^2_{31}$ relative to the value obtained in the perfectly-coherent hypothesis. 
When $\sigma$ is large enough, decoherence effects are outside the reach of Daya Bay and RENO and hence the horizontal allowed regions in Fig.~\ref{fig:sigma_X} (left) and Fig.~\ref{fig:sigma_X} (right) extend to arbitrarily large $\sigma$.

\begin{figure}[t]
\centering
\includegraphics[width=0.49\textwidth]{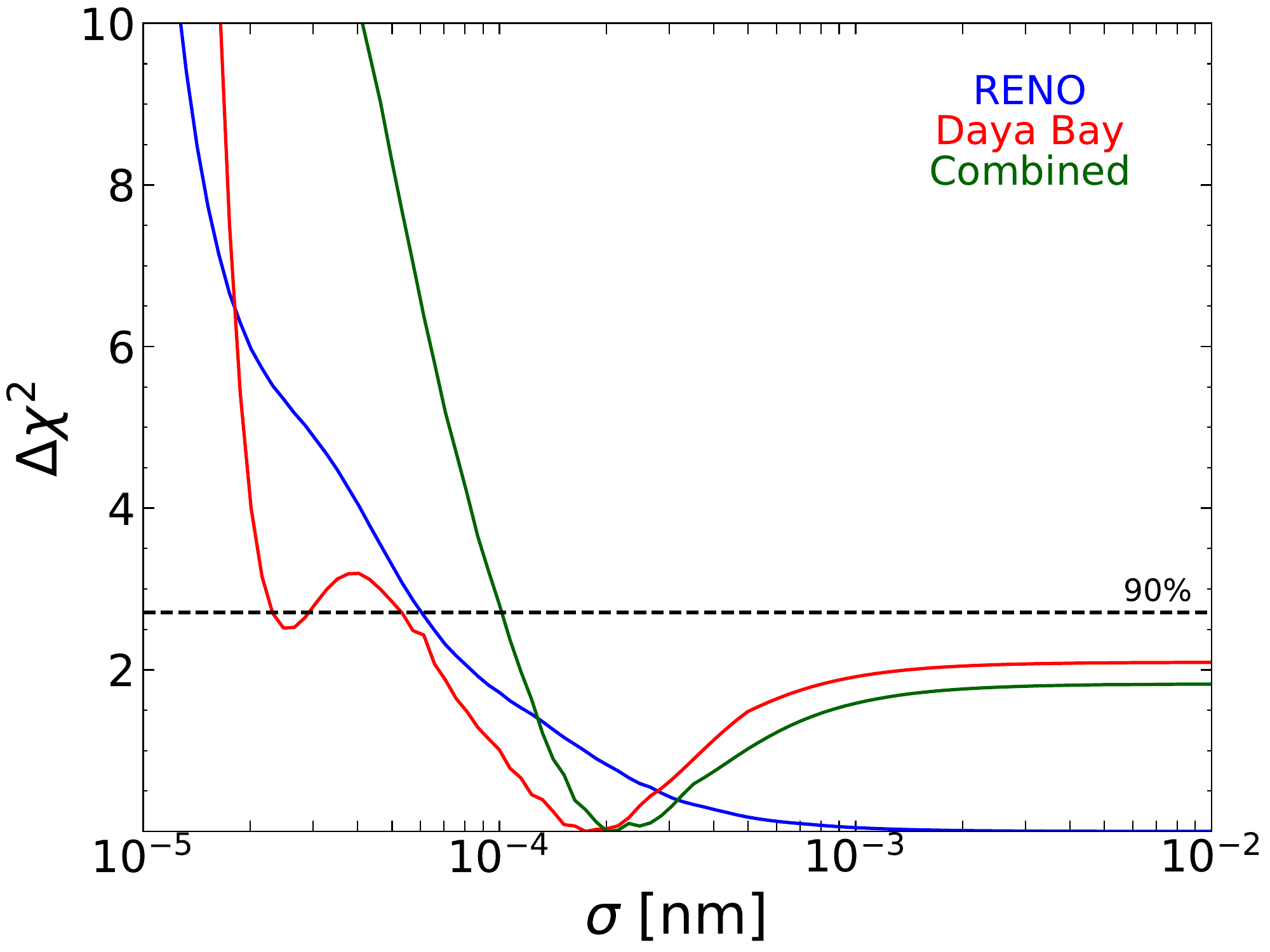}
\caption{
The reduced $\chi^2$ as a function of $\sigma$ relative to its minimum value, obtained from the analysis of RENO (blue), Daya Bay (red) and from the combined analysis of both experiments (green).}
\label{fig:sigma_profile}
\end{figure}

Marginalizing over $\Delta m^2_{31}$ and $\sin^2\theta_{13}$, we extract the reduced $\chi^2(\sigma)$, depicted relative to its minimum value in Fig.~\ref{fig:sigma_profile}; the minimum corresponds to $\sigma = 2.01\times10^{-4}$~nm. 
Arbitrarily large values of $\sigma$ are allowed at better than the 90\% CL and we translate the information in Fig.~\ref{fig:sigma_profile} into the lower bound $\sigma > 1.02\times10^{-4}$~nm at 90\% CL,  combining data from RENO and Daya Bay.
For $E=3$~MeV and $\Delta m^2_{31}=2.5\times 10^{-3}$~eV$^2$, this translates into $L^{\rm coh}_{13}>1.8$~km. 
This is consistent with the naive expectation that RENO and Daya Bay should be sensitive to $L^{\rm coh}_{13}\lesssim{\cal O}(1~\rm km)$.

\section{Sensitivity of the JUNO Experimental Setup}
\label{sec:JUNO}

In this section we study the sensitivity of the future JUNO experiment~\cite{An:2015jdp} to constrain or measure the neutrino wave-packet width $\sigma$.
We first estimate the sensitivity of JUNO to $\sigma$ assuming future JUNO data are consistent with no decoherence effects, $\sigma \to \infty$.
Next, we check the potential of JUNO to establish and measure the presence of decoherence assuming the future JUNO data are consistent with $\sigma = 2.01\times10^{-4}$~nm, the best-fit value of $\sigma$ from current reactor data, discussed in the previous section. 

In order to simulate JUNO data, we make use of information from Ref.~\cite{Bezerra:2019dao}. In particular, we assume the 10-reactor configuration. Thermal powers and baselines can be found in Ref.~\cite{An:2015jdp} while fluxes, cross sections, and fission fractions are fixed to the ones we used in our analyses of Daya Bay data. When computing oscillation probabilities, we ignore matter effects, which are subdominant. For more details, we refer readers to Refs.~\cite{Li:2016txk,Khan:2019doq}. There, it was demonstrated that, when pursuing oscillation analyses, matter effects primarily impact, very slightly, the extraction of best-fit values but are negligible when it comes to uncertainties and the sensitivity to other effects, including the mass ordering~\cite{Li:2016txk,Khan:2019doq}. 
Our statistical analyses are performed with

\begin{equation}
 \chi^2_\text{JUNO}(\vec{p}) = \min_{\vec{\alpha}}\left\{\sum_{i=1}^{N_\text{JUNO}}\left(\frac{N_{\text{dat},i} - N_{\text{exp},i}(\vec{p},\vec{\alpha})}{\sigma^\text{JUNO}_i}\right)^2 + \sum_k \left(\frac{\alpha_k - \mu_k}{\sigma_k} \right)^2\right\}\,.
\end{equation}
We assume JUNO will run for 6 years, corresponding to 1800 days of data taking~\cite{Bezerra:2019dao} and we do not assume the existence of a near detector. 
The systematic uncertainties are virtually the same as the ones discussed in the last section but, in order to account for the absence of a near detector, we include an overall flux-normalization uncertainty due to unknowns in the reactor flux spectrum. 

When simulating data, unless otherwise noted, we assume the true values of the oscillation parameters to be those spelled out in Eq.~(\ref{eq:params}) and, as discussed earlier, assume the mass-ordering is known to be normal. We expect very similar results if it turns out that the mass-ordering is known to be inverted when JUNO takes data. Furthermore, since the impact of the mass-ordering on JUNO data is very different from the effects of non-trivial decoherence, we also expect similar results if one were to assume, in the data analysis, that the mass-ordering is not known. We do not pursue this line of investigation further as it combines different goals of JUNO in a complicated, and not especially illuminating, way. One of the main goals of JUNO is to determine the neutrino mass-ordering by performing an exquisite measurement of the oscillation probability as a function of energy with a baseline that is long enough so both $\Delta m^2_{21}$ and $\Delta m^2_{31}$ effects can be observed. Allowing for the hypothesis that $\sigma$ is finite will, of course, render such an analysis more challenging. Determining how much more challenging is outside the aspirations of this manuscript.

\subsection{Ruling Out Decoherence}

\begin{figure}[t]
\centering
\includegraphics[width=0.49\textwidth]{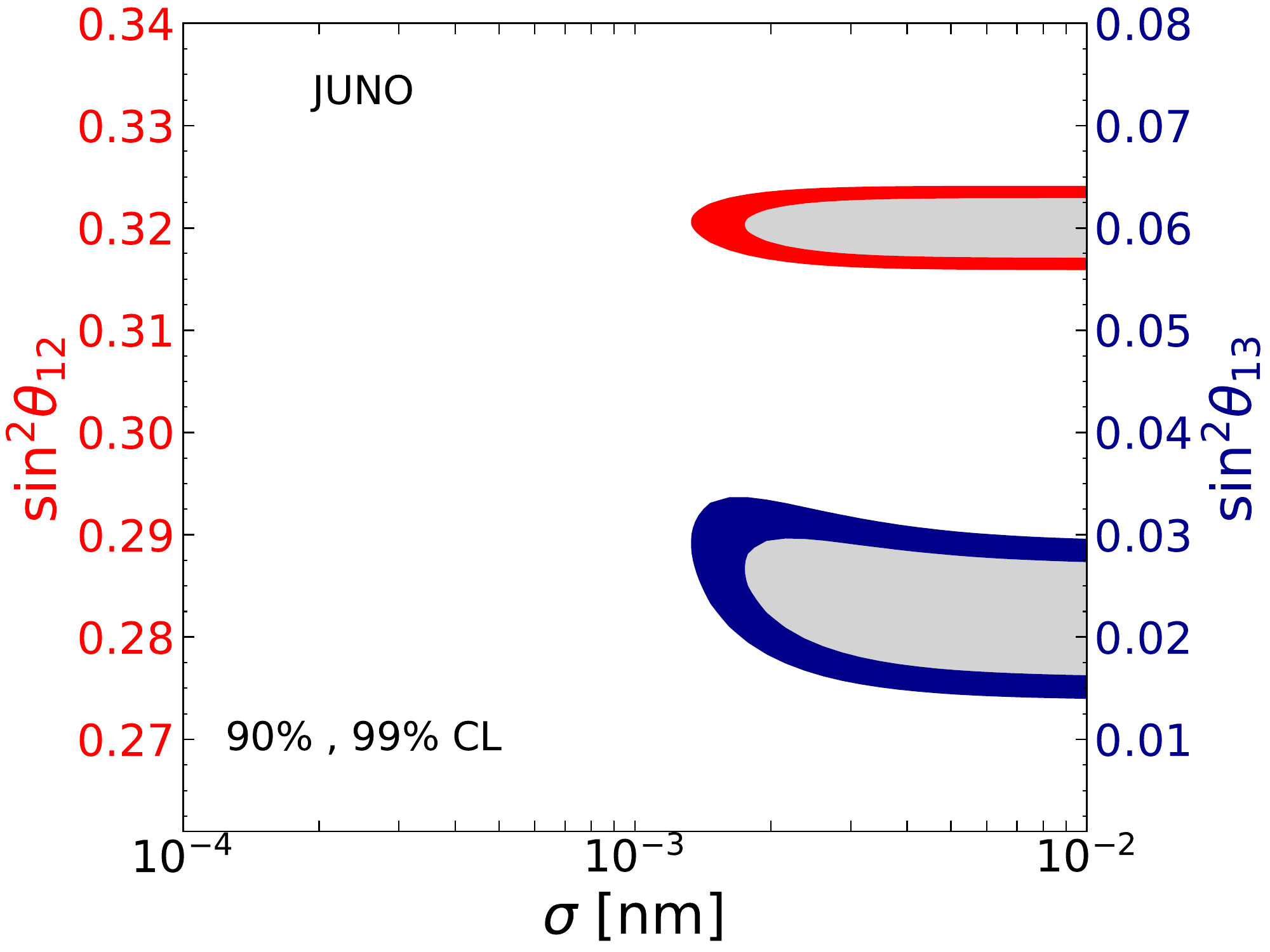}
\includegraphics[width=0.49\textwidth]{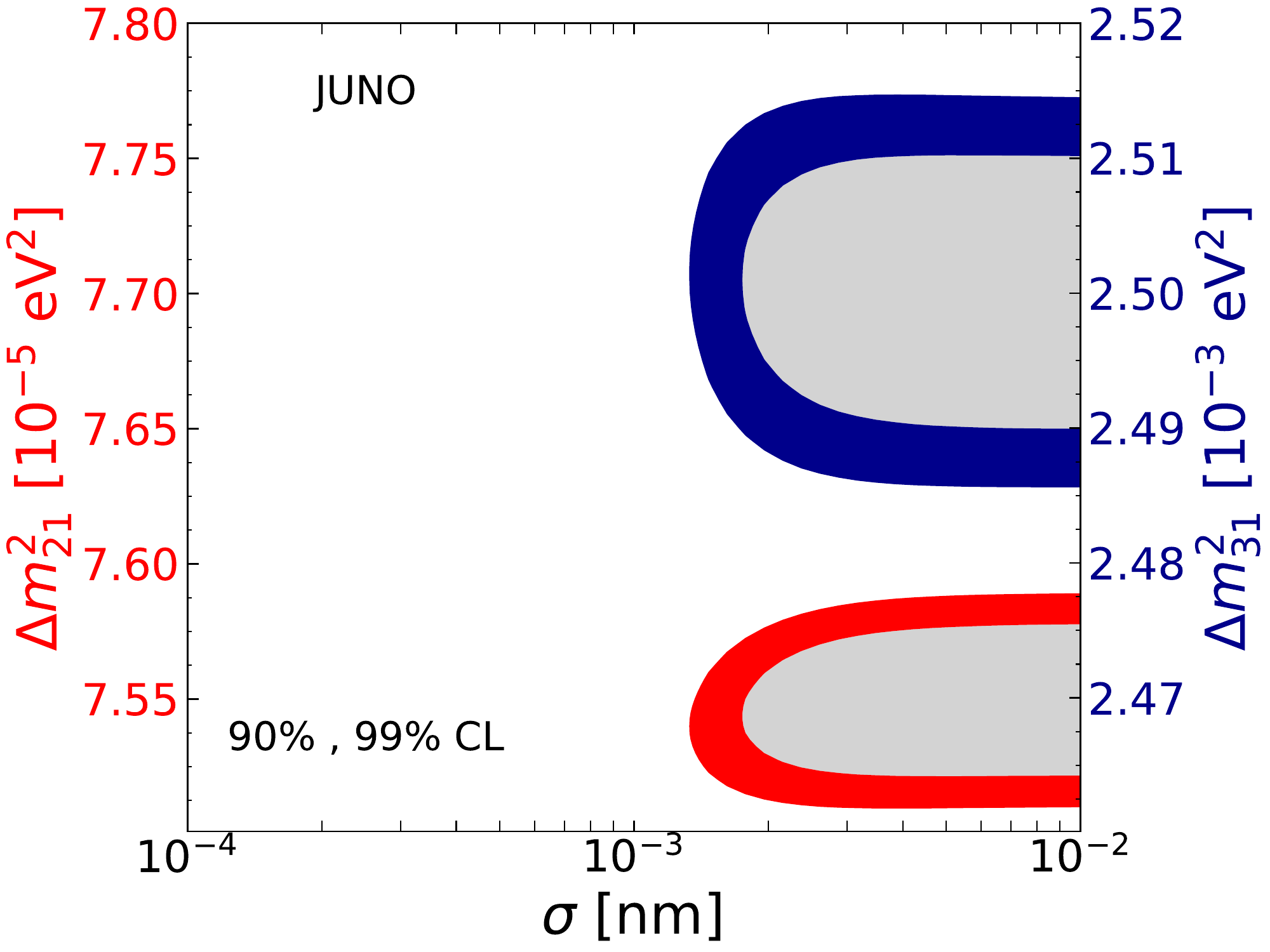}
\caption{
90 and 99\% CL (2 d.o.f.) allowed regions in the $\sigma$--mixing angle (left) and $\sigma$--$\Delta m^2$ planes obtained by analyzing simulated JUNO data consistent with no decoherence ($\sigma\to\infty$) . The red (blue) contours have to be compared with the red (blue) axes in both panels. The input values for the standard oscillation parameters are listed in Eqs.~(\ref{eq:params}).}
\label{fig:JUNO_sigma_sq1X}
\end{figure}

\begin{figure}[t]
\centering
\includegraphics[width=0.49\textwidth]{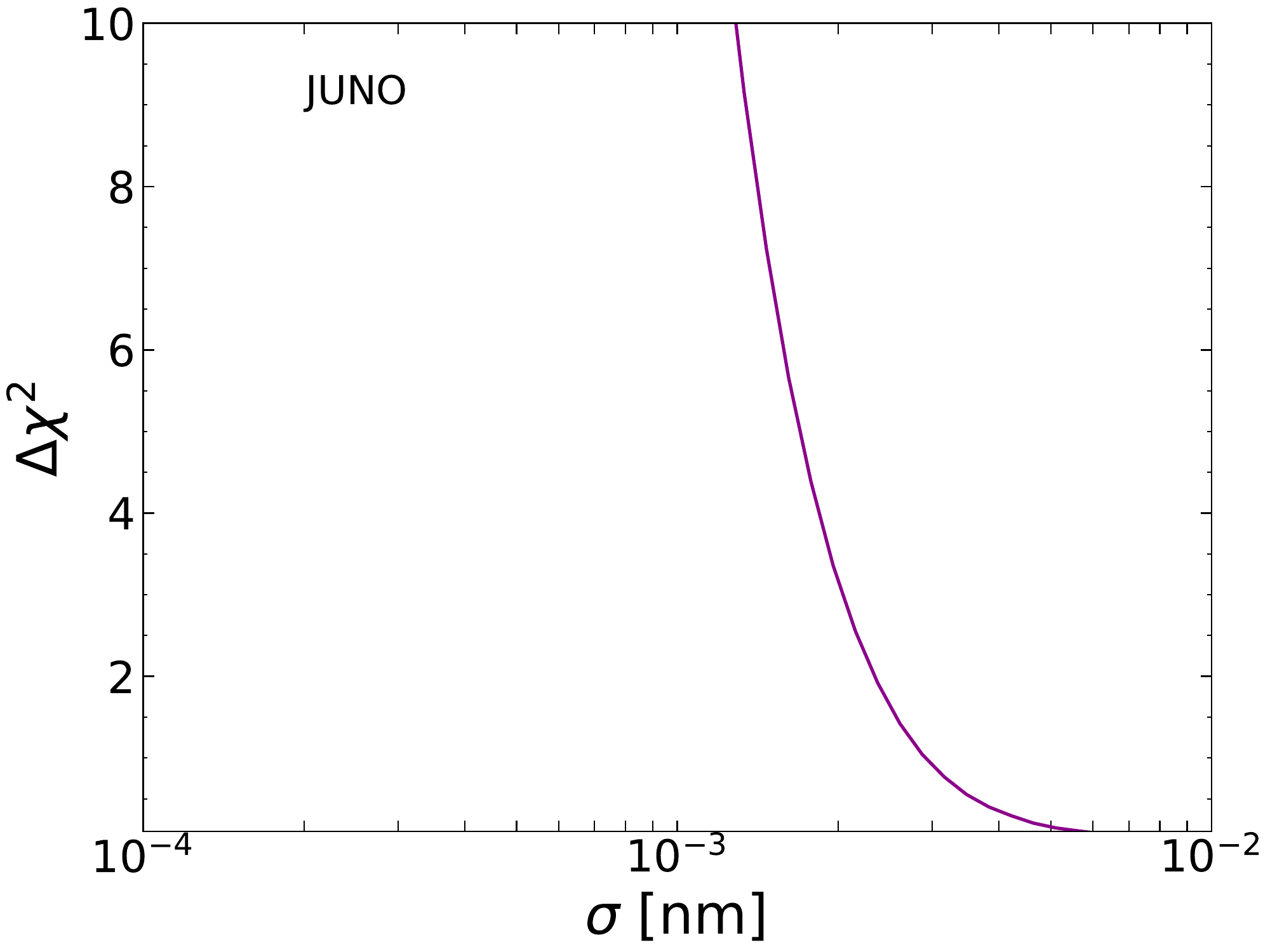}
\caption{
The reduced $\chi^2$ as a function of $\sigma$ relative to its minimum value, obtained by analyzing simulated JUNO data consistent with no decoherence ($\sigma\to\infty$) and marginalizing over the remaining four neutrino oscillation parameters. The input values for the standard oscillation parameters are listed in Eqs.~(\ref{eq:params}).}
\label{fig:JUNO_sigma_profile}
\end{figure}

Here, we simulate data consistent with no decoherence ($\sigma\to\infty$) and analyze them as discussed above. Fig.~\ref{fig:JUNO_sigma_sq1X} depicts the allowed regions of the $\sigma$--mixing angles (left) and $\sigma$--$\Delta m^2$'s (right) parameter spaces. When generating these two-dimensional regions, we marginalize over all absent parameters. 
Fig.~\ref{fig:JUNO_sigma_sq1X} reveals that the precision with which JUNO can measure the different oscillation parameters is not significantly impacted by allowing for the possibility that $\sigma$ is finite. The reason for this is that JUNO is sensitive to several oscillation maxima and minima associated to the short oscillation lengths and the degeneracies observed in Daya Bay and RENO are completely lifted. Furthermore, the absence of decoherence effects associated with $L^{\rm coh}_{13}$ and $L^{\rm coh}_{23}$ preclude observable $L^{\rm coh}_{12}$ effects since $L^{\rm coh}_{12}/L^{\rm coh}_{13}\sim 30$ and $\sin^2\theta_{13}$ effects are clearly visible.
Fig.~\ref{fig:JUNO_sigma_profile} depicts the reduced $\chi^2(\sigma)$, obtained upon marginalizing over all four oscillation parameters. These data would translate into $\sigma > 2.11\times 10^{-3}$~nm at the 90\% CL. This is more than a factor of 20 stronger than the current bound from RENO and Daya Bay, obtained in the last section. 
\subsection{Observing and Measuring Decoherence}

Here, we simulate data consistent with the solar parameters from Eq.~(\ref{eq:params}) and the best-fit value obtained from the analysis of Daya Bay and RENO data performed in the last section. For the decoherence parameter we set $\sigma = 2.01\times10^{-4}$~nm, while for the standard neutrino oscillation parameters we have $\Delta m_{31}^2 = 2.63\times 10^{-3}$~eV$^2$ and $\sin^2\theta_{13} = 0.0231$.
Fig.~\ref{fig:osc_prob} reveals that the impact of decoherence is very strong in JUNO, and we expect the no-decoherence hypothesis to be completely ruled out. Furthermore, the short-wavelength oscillations are completely erased, rendering the measurements of $\Delta m^2_{31}$ and $\Delta m^2_{32}$ impossible. It is very clear that, under these circumstances, JUNO is completely insensitive to the mass ordering.\footnote{Qualitatively, we estimate that capability of JUNO to determine the mass ordering is negatively impacted by decoherence effects if $\sigma$ turns out to be, roughly, a few times $10^{-3}$~nm or smaller. For these values, decoherence effects are non-trivial in JUNO, see Fig.~\ref{fig:JUNO_sigma_sq1X}.}

\begin{figure}[t]
\centering
\includegraphics[width=0.8\textwidth]{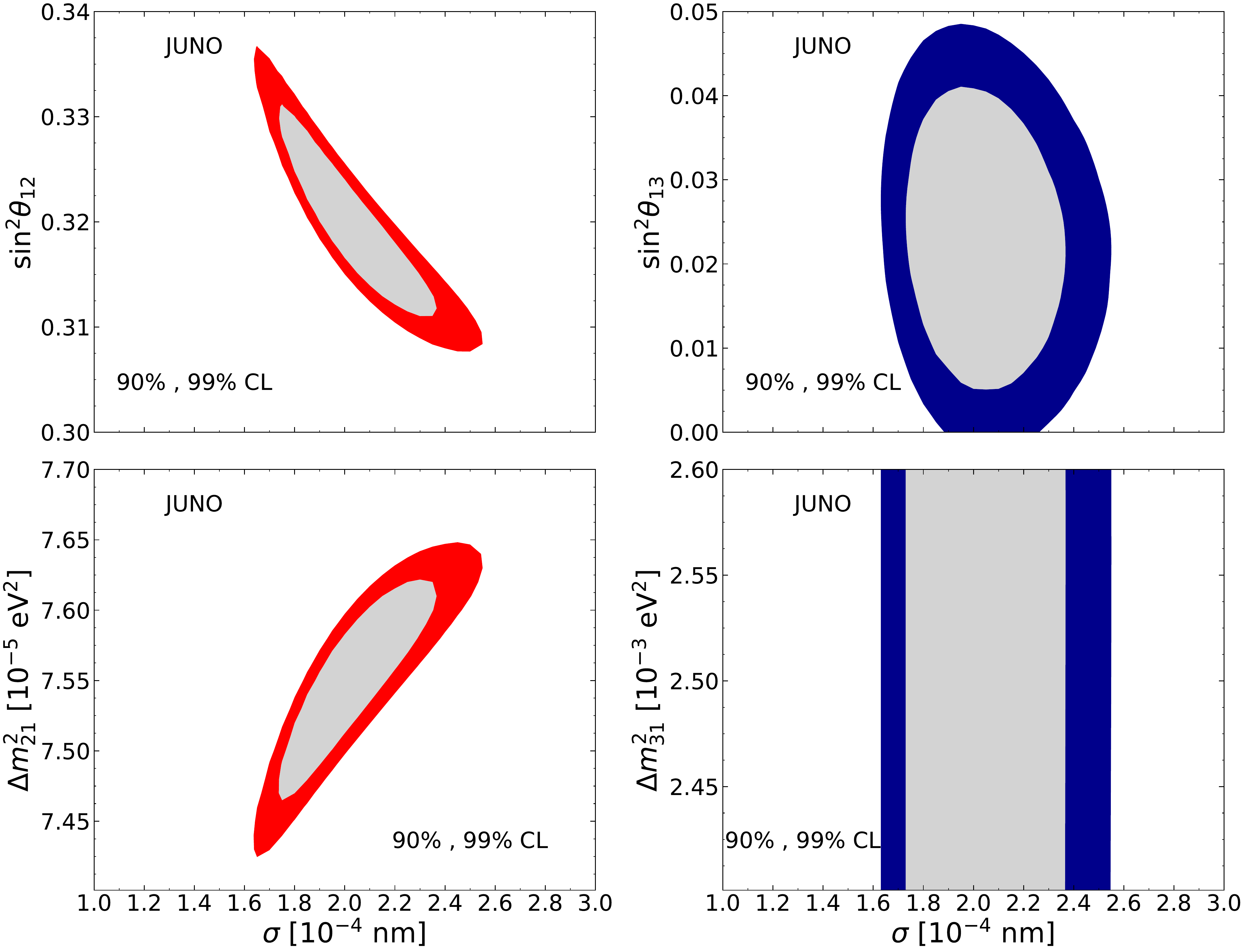}
\caption{
90 and 99\% CL (2 d.o.f.) allowed regions in the $\sigma$--$\sin^2\theta_{12}$ (top,left),  $\sigma$--$\Delta m^2_{21}$ (bottom,left),  $\sigma$--$\sin^2\theta_{13}$ (top,right), and  $\sigma$--$\Delta m^2_{31}$ (bottom, right) planes, obtained by analyzing simulated JUNO data consistent with strong decoherence ($\sigma=2.01\times 10^{-4}$~nm). The input values for the standard solar oscillation parameters are listed in Eqs.~(\ref{eq:params}) while $\Delta m_{31}^2 = 2.63\times 10^{-3}$~eV$^2$ and $\sin^2\theta_{13} = 0.0231$. See text for details.}
\label{fig:JUNO_dec_planes}
\end{figure}

Fig.~\ref{fig:JUNO_dec_planes} depicts the allowed regions of the $\sigma$--mixing angles ($\sin^2\theta_{12}$ on the top, left and $\sin^2\theta_{13}$ on the top, right) and $\sigma$--$\Delta m^2$'s ($\Delta m^2_{21}$ on the bottom, left and $\Delta m^2_{31}$ on the bottom, right)  parameter spaces. 
When generating all two-dimensional regions, we marginalize over all absent parameters. 
As advertised, there is no sensitivity to $\Delta m^2_{31}$ (Fig.~\ref{fig:JUNO_dec_planes} [bottom, right]). %
Nonetheless, averaged-out effects of the short-wavelength oscillations remain and one can measure $\sin^2\theta_{13}$ with finite, albeit poorer, precision, since now $\sin^2\theta_{13} = 0$ remains allowed at 99\% CL (cf. Fig.~\ref{fig:JUNO_sigma_sq1X} [left]). Long-wavelength effects are still present and hence both $\Delta m^2_{21}$ and $\sin^2\theta_{12}$ can be measured, see Fig.~\ref{fig:JUNO_dec_planes} (left). 
Similar to what we observe for RENO and Daya Bay, measurements of the oscillation frequency and amplitude are strongly correlated with those of $\sigma$. 
Smaller sigma translate into larger $\sin^2\theta_{12}$ in order to compensate for the flattened-out oscillation probability while smaller sigma translate into smaller $\Delta m^2_{21}$ in order to compensate for the shift of the oscillation maximum to larger energies (smaller $L/E$). %
These degeneracies lead to a less precise determination of the solar parameters (cf. Fig.~\ref{fig:JUNO_sigma_sq1X}).

\begin{figure}[!htb]
\centering
\includegraphics[width=0.49\textwidth]{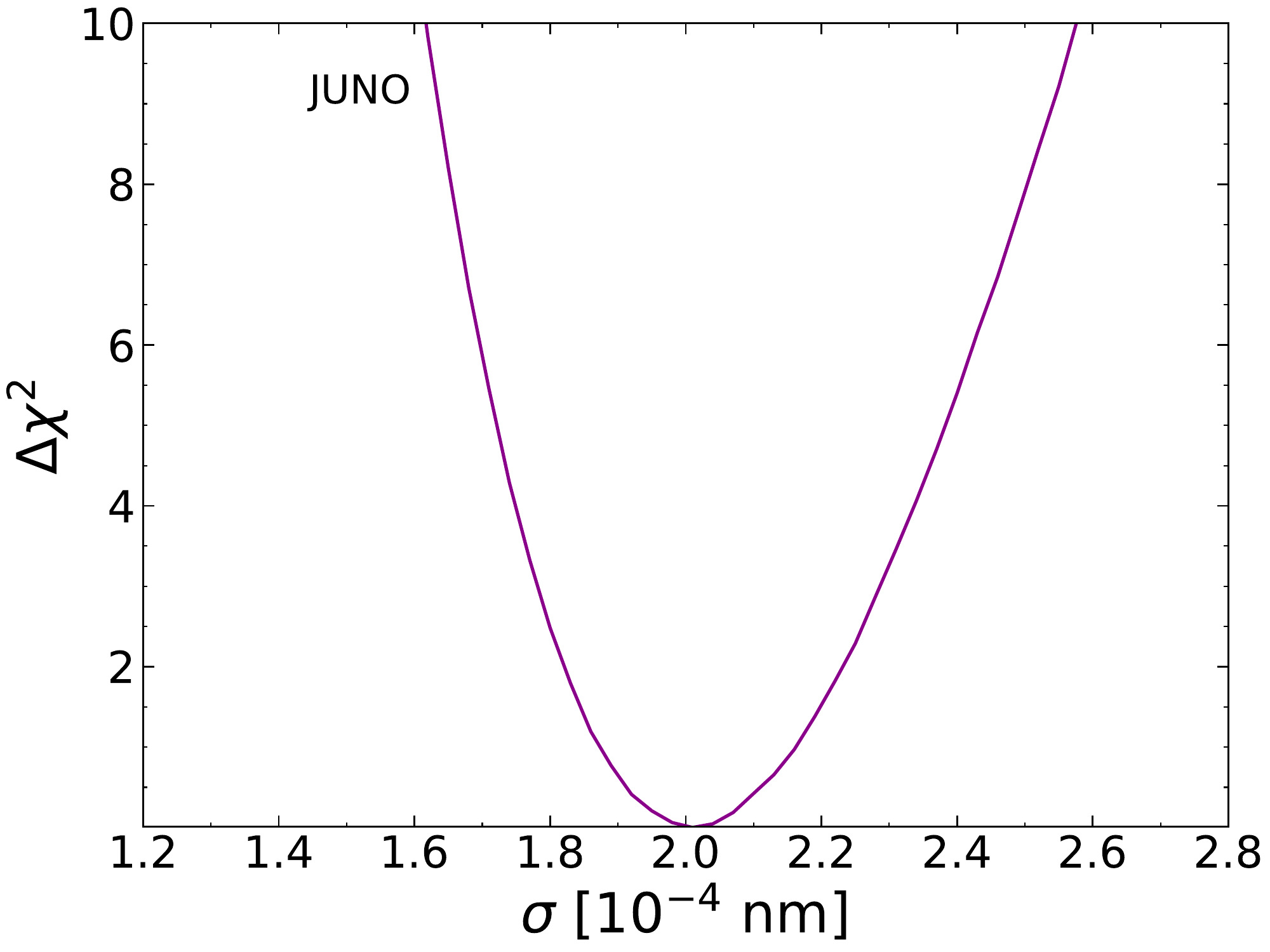}
\caption{
The reduced $\chi^2$ as a function of $\sigma$ relative to its minimum value,  obtained by analyzing simulated JUNO data consistent with strong decoherence ($\sigma=2.01\times 10^{-4}$~nm) and marginalizing over all oscillation parameters. The input values for the standard solar oscillation parameters are listed in Eqs.~(\ref{eq:params}) while $\Delta m_{31}^2 = 2.63\times 10^{-3}$~eV$^2$ and $\sin^2\theta_{13} = 0.0231$. See text for details.}
\label{fig:JUNO_dec_profile}
\end{figure}

Fig.~\ref{fig:JUNO_dec_profile} depicts the reduced $\chi^2(\sigma)$, relative to the minimum value. 
A clear measurement of the neutrino-wave-packet width can be extracted: $\sigma=\left(2.01_{-0.14}^{+0.16}\right)\times 10^{-4}$~nm. The no-decoherence hypothesis is ruled out at more than ten~$\sigma$.
%

\section{Conclusions}
\label{sec:conc}

Neutrinos observed in all neutrino oscillation experiments, to date, can either be treated as perfectly incoherent -- e.g., solar neutrino experiments modulo Earth matter-effects -- or perfectly coherent -- e.g., Daya Bay and RENO -- superpositions of the mass eigenstates. 
The position-dependent loss of coherence expected, in principle, of neutrinos produced and detected under any circumstances, has never been observed. 

Here, we explore how well reactor antineutrino experiments can constrain or measure the loss of coherence of reactor antineutrinos. 
For concreteness, we assume that decoherence effects are captured by the size of the neutrino wave-packet, $\sigma$. 
A perfectly coherent neutrino beam corresponds to $\sigma\to\infty$ while an incoherent superposition of mass eigenstates is associated to $\sigma=0$. 
We expect reactor neutrino experiments to be excellent laboratories to study decoherence given the high statistics, the compactness of sources and detectors, including good position resolution, excellent event-by-event energy reconstruction, and very long baselines. 

We find that current reactor data from Daya Bay and RENO constrain $\sigma>1.0\times 10^{-4}$~nm while future data from JUNO should be sensitive to $\sigma<2.1\times 10^{-3}$~nm, a factor of 20 more sensitive than the current data. 
If $\sigma\sim {\rm few}\times10^{-4}$~nm, in perfect agreement with current reactor neutrino data, we expect decoherence effects to be clearly visible in JUNO, as illustrated in Fig.~\ref{fig:osc_prob}. In this case, $\sigma$ should be measured in JUNO with good precision. 

One can naively estimate that, for neutrinos produced in nuclear reactors and detected via inverse beta-decay, $\sigma$ should be, at least, of order of the typical interatomic spacing that characterizes the fuel inside the nuclear reactor, which we anticipate is safely outside the sensitivity of JUNO.  For the sake of reference, for pure, solid uranium, lattice parameters are of order 0.1--1~nm. JUNO is, however, sensitive to other distance scales associated with electron antineutrinos from beta-decay, including the typical size of the beta-decaying nuclei -- around $10^{-5}$~nm -- or the inverse of the neutrino energy, $1/E\sim 10^{-4}$~nm. The discovery of nontrivial decoherence effects in JUNO would indicate that our understanding of the coherence of neutrino sources (or quantum mechanics?) is, at least, incomplete.

Outside of the decoherence effects discussed here, other new phenomena can impact the survival probability of reactor antineutrinos, including very fast neutrino decay into lighter neutrinos or new, very light particles \cite{Abrahao:2015rba,Porto-Silva:2020gma} and a variety of new-physics effects~
\cite{Khan:2013hva,Ohlsson:2013nna,Bakhti:2014pva,Li:2014rya,Liao:2017awz,Anamiati:2019maf}.
These new-physics effects modify the survival probability in a way that is qualitatively different from the decoherence effects discussed here so we do not expect, assuming the data are not consistent with the standard three-neutrino paradigm, that it would be difficult to distinguish strong decoherence in neutrino propagation from other new physics. A quantitative study of how well one can distinguish different, new phenomena with JUNO is outside the scope of this manuscript. 

\section*{Acknowledgments}
We would like to thank David Vanegas Forero and Pablo Mart\'inez Mirav\'e for useful discussions, and Roni Harnik, Boris Kayser, and Pedro Machado for inspiring conversations. We are very grateful to Soo-Bong Kim from the RENO collaboration for providing the digitized data for 2900 days of running time. VDR and CAT are also very grateful for the kind hospitality received at Fermilab where this work was initiated.
This work was supported by the Spanish grants FPA2017-90566-REDC (Red Consolider MultiDark), FPA2017-85216-P and PROMETEO/2018/165 (Generalitat Valenciana). VDR acknowledges financial support by the SEJI/2018/033 grant, funded by Generalitat Valenciana. CAT is supported by the FPI grant BES-2015-073593. The work of AdG is supported in part by the DOE Office of Science award \#DE-SC0010143. 
\bibliographystyle{kp}

\begin{thebibliography}{50}
\expandafter\ifx\csname natexlab\endcsname\relax\def\natexlab#1{#1}\fi

\bibitem[Nussinov(1976)]{Nussinov:1976uw}
S.~Nussinov, ``{Solar Neutrinos and Neutrino Mixing}'', {\em Phys. Lett. B}
  {\bfseries 63} (1976) 201--203.

\bibitem[Kayser(1981)]{Kayser:1981ye}
B.~Kayser, ``{On the Quantum Mechanics of Neutrino Oscillation}'', {\em Phys.
  Rev. D} {\bfseries 24} (1981) 110.

\bibitem[Giunti et~al.(1991)Giunti, Kim, and Lee]{Giunti:1991ca}
C.~Giunti, C.~Kim, and U.~Lee, ``{When do neutrinos really oscillate?: Quantum
  mechanics of neutrino oscillations}'', {\em Phys. Rev. D} {\bfseries 44}
  (1991) 3635--3640.

\bibitem[Giunti et~al.(1992)Giunti, Kim, and Lee]{Giunti:1991sx}
C.~Giunti, C.~Kim, and U.~Lee, ``{Coherence of neutrino oscillations in vacuum
  and matter in the wave packet treatment}'', {\em Phys.\ Lett.\ B} {\bfseries
  274} (1992) 87--94.

\bibitem[Kiers and Weiss(1998)]{Kiers:1997pe}
K.~Kiers and N.~Weiss, ``{Neutrino oscillations in a model with a source and
  detector}'', {\em Phys. Rev. D} {\bfseries 57} (1998) 3091--3105,
  \href{https://arxiv.org/abs/hep-ph/9710289}{{\ttfamily hep-ph/9710289}}.

\bibitem[Giunti and Kim(1998)]{Giunti:1997wq}
C.~Giunti and C.~Kim, ``{Coherence of neutrino oscillations in the wave packet
  approach}'', {\em Phys. Rev. D} {\bfseries 58} (1998) 017301,
  \href{https://arxiv.org/abs/hep-ph/9711363}{{\ttfamily hep-ph/9711363}}.

\bibitem[Grimus et~al.(1999)Grimus, Stockinger, and Mohanty]{Grimus:1998uh}
W.~Grimus, P.~Stockinger, and S.~Mohanty, ``{The Field theoretical approach to
  coherence in neutrino oscillations}'', {\em Phys. Rev. D} {\bfseries 59}
  (1999) 013011,  \href{https://arxiv.org/abs/hep-ph/9807442}{{\ttfamily
  hep-ph/9807442}}.

\bibitem[Akhmedov and Kopp(2010)]{Akhmedov:2010ms}
E.~K. Akhmedov and J.~Kopp, ``{Neutrino Oscillations: Quantum Mechanics vs.
  Quantum Field Theory}'', {\em JHEP} {\bfseries 04} (2010) 008,
  \href{https://arxiv.org/abs/1001.4815}{{\ttfamily arXiv:1001.4815}},
  [Erratum: JHEP 10, 052 (2013)].

\bibitem[Naumov and Naumov(2010)]{Naumov:2010um}
D.~Naumov and V.~Naumov, ``{A Diagrammatic treatment of neutrino
  oscillations}'', {\em J. Phys. G} {\bfseries 37} (2010) 105014,
  \href{https://arxiv.org/abs/1008.0306}{{\ttfamily arXiv:1008.0306}}.

\bibitem[Akhmedov et~al.(2012)Akhmedov, Hernandez, and
  Smirnov]{Akhmedov:2012uu}
E.~Akhmedov, D.~Hernandez, and A.~Smirnov, ``{Neutrino production coherence and
  oscillation experiments}'', {\em JHEP} {\bfseries 04} (2012) 052,
  \href{https://arxiv.org/abs/1201.4128}{{\ttfamily arXiv:1201.4128}}.

\bibitem[de~Salas et~al.(2018)de~Salas, Forero, Ternes, T\'ortola, and
  Valle]{deSalas:2017kay}
P.~F. de~Salas, D.~V. Forero, C.~A. Ternes, M.~T\'ortola, and J.~W.~F. Valle,
  ``{Status of neutrino oscillations 2018: 3$\sigma$ hint for normal mass
  ordering and improved CP sensitivity}'', {\em Phys. Lett.} {\bfseries B782}
  (2018) 633--640,
 \href{https://arxiv.org/abs/1708.01186}{{\ttfamily arXiv:1708.01186}}.

\bibitem[Kiers et~al.(1996)Kiers, Nussinov, and Weiss]{Kiers:1995zj}
K.~Kiers, S.~Nussinov, and N.~Weiss, ``{Coherence effects in neutrino
  oscillations}'', {\em Phys. Rev. D} {\bfseries 53} (1996) 537--547,
  \href{https://arxiv.org/abs/hep-ph/9506271}{{\ttfamily hep-ph/9506271}}.

\bibitem[Ohlsson(2001)]{Ohlsson:2000mj}
T.~Ohlsson, ``{Equivalence between neutrino oscillations and neutrino
  decoherence}'', {\em Phys. Lett. B} {\bfseries 502} (2001) 159--166,
  \href{https://arxiv.org/abs/hep-ph/0012272}{{\ttfamily hep-ph/0012272}}.

\bibitem[Beuthe(2003)]{Beuthe:2001rc}
M.~Beuthe, ``{Oscillations of neutrinos and mesons in quantum field theory}'',
  {\em Phys. Rept.} {\bfseries 375} (2003) 105--218,
  \href{https://arxiv.org/abs/hep-ph/0109119}{{\ttfamily hep-ph/0109119}}.

\bibitem[Beuthe(2002)]{Beuthe:2002ej}
M.~Beuthe, ``{Towards a unique formula for neutrino oscillations in vacuum}'',
  {\em Phys.\ Rev.\ D} {\bfseries 66} (2002) 013003,
  \href{https://arxiv.org/abs/hep-ph/0202068}{{\ttfamily hep-ph/0202068}}.

\bibitem[Giunti(2004)]{Giunti:2003ax}
C.~Giunti, ``{Coherence and wave packets in neutrino oscillations}'', {\em
  Found. Phys. Lett.} {\bfseries 17} (2004) 103--124,
  \href{https://arxiv.org/abs/hep-ph/0302026}{{\ttfamily hep-ph/0302026}}.

\bibitem[Blennow et~al.(2005)Blennow, Ohlsson, and Winter]{Blennow:2005yk}
M.~Blennow, T.~Ohlsson, and W.~Winter, ``{Damping signatures in future neutrino
  oscillation experiments}'', {\em JHEP} {\bfseries 06} (2005) 049,
  \href{https://arxiv.org/abs/hep-ph/0502147}{{\ttfamily hep-ph/0502147}}.

\bibitem[Farzan and Smirnov(2008)]{Farzan:2008eg}
Y.~Farzan and A.~Y. Smirnov, ``{Coherence and oscillations of cosmic
  neutrinos}'', {\em Nucl. Phys. B} {\bfseries 805} (2008) 356--376,
  \href{https://arxiv.org/abs/0803.0495}{{\ttfamily arXiv:0803.0495}}.

\bibitem[Kayser and Kopp(2010)]{Kayser:2010pr}
B.~Kayser and J.~Kopp, ``{Testing the Wave Packet Approach to Neutrino
  Oscillations in Future Experiments}'',
  \href{https://arxiv.org/abs/1005.4081}{{\ttfamily arXiv:1005.4081}}.

\bibitem[Naumov(2013)]{Naumov:2013uia}
D.~Naumov, ``{On the Theory of Wave Packets}'', {\em Phys. Part. Nucl. Lett.}
  {\bfseries 10} (2013) 642--650,
  \href{https://arxiv.org/abs/1309.1717}{{\ttfamily arXiv:1309.1717}}.

\bibitem[Jones(2015)]{Jones:2014sfa}
B.~Jones, ``{Dynamical pion collapse and the coherence of conventional neutrino
  beams}'', {\em Phys. Rev. D} {\bfseries 91} (2015), no.~5, 053002,
  \href{https://arxiv.org/abs/1412.2264}{{\ttfamily arXiv:1412.2264}}.

\bibitem[Akhmedov(2019)]{Akhmedov:2019iyt}
E.~Akhmedov, ``{Quantum mechanics aspects and subtleties of neutrino
  oscillations}'', in ``{International Conference on History of the Neutrino}:
  {1930-2018}''.
\newblock 2019.
\newblock  \href{https://arxiv.org/abs/1901.05232}{{\ttfamily
  arXiv:1901.05232}}.

\bibitem[Grimus(2019)]{Grimus:2019hlq}
W.~Grimus, ``{Revisiting the quantum field theory of neutrino oscillations in
  vacuum}'',  \href{https://arxiv.org/abs/1910.13776}{{\ttfamily
  arXiv:1910.13776}}.

\bibitem[Naumov and Naumov(2020)]{Naumov:2020yyv}
D.~Naumov and V.~Naumov, ``{Quantum Field Theory of Neutrino Oscillations}'',
  {\em Phys. Part. Nucl.} {\bfseries 51} (2020), no.~1, 1--106.

\bibitem[An et~al.(2017)]{An:2016pvi}
{\bfseries Daya Bay} Collaboration, F.~P. An {\em et~al.}, ``{Study of the wave
  packet treatment of neutrino oscillation at Daya Bay}'', {\em Eur.\ Phys.\
  J.\ C} {\bfseries 77} (2017), no.~9, 606,
  \href{https://arxiv.org/abs/1608.01661}{{\ttfamily arXiv:1608.01661}}.

\bibitem[Yoo(2020)]{RENO-neutrino2020}
{\bfseries RENO} Collaboration, J.~Yoo, ``Recent results from reno
  experiment''.
  \url{https://indico.fnal.gov/event/43209/contributions/187886/attachments/130339/158753/Neutrino2020YooRENO.pdf},
  July 2020.

\bibitem[Adey et~al.(2018)]{Adey:2018zwh}
{\bfseries Daya Bay} Collaboration, D.~Adey {\em et~al.}, ``{Measurement of the
  Electron Antineutrino Oscillation with 1958 Days of Operation at Daya Bay}'',
  {\em Phys. Rev. Lett.} {\bfseries 121} (2018), no.~24, 241805,
 \href{https://arxiv.org/abs/1809.02261}{{\ttfamily arXiv:1809.02261}}.

\bibitem[Ahn et~al.(2010)]{Ahn:2010vy}
{\bfseries RENO} Collaboration, J.~Ahn {\em et~al.}, ``{RENO: An Experiment for
  Neutrino Oscillation Parameter $\theta_{13}$ Using Reactor Neutrinos at
  Yonggwang}'',  \href{https://arxiv.org/abs/1003.1391}{{\ttfamily
  arXiv:1003.1391}}.

\bibitem[Seo et~al.(2018)]{Seo:2016uom}
{\bfseries RENO} Collaboration, S.~Seo {\em et~al.}, ``{Spectral Measurement of
  the Electron Antineutrino Oscillation Amplitude and Frequency using 500 Live
  Days of RENO Data}'', {\em Phys.\ Rev.\ D} {\bfseries 98} (2018), no.~1,
  012002,  \href{https://arxiv.org/abs/1610.04326}{{\ttfamily
  arXiv:1610.04326}}.

\bibitem[Bak et~al.(2018)]{Bak:2018ydk}
{\bfseries RENO} Collaboration, G.~Bak {\em et~al.}, ``{Measurement of Reactor
  Antineutrino Oscillation Amplitude and Frequency at RENO}'', {\em Phys. Rev.
  Lett.} {\bfseries 121} (2018), no.~20, 201801,
 \href{https://arxiv.org/abs/1806.00248}{{\ttfamily arXiv:1806.00248}}.

\bibitem[An et~al.(2017{\natexlab{a}})]{An:2016srz}
{\bfseries Daya Bay} Collaboration, F.~P. An {\em et~al.}, ``{Improved
  Measurement of the Reactor Antineutrino Flux and Spectrum at Daya Bay}'',
  {\em Chin. Phys.} {\bfseries C41} (2017){\natexlab{a}}, no.~1, 013002,
 \href{https://arxiv.org/abs/1607.05378}{{\ttfamily arXiv:1607.05378}}.

\bibitem[An et~al.(2017{\natexlab{b}})]{An:2016ses}
{\bfseries Daya Bay} Collaboration, F.~P. An {\em et~al.}, ``{Measurement of
  electron antineutrino oscillation based on 1230 days of operation of the Daya
  Bay experiment}'', {\em Phys. Rev.} {\bfseries D95} (2017){\natexlab{b}},
  no.~7, 072006,
 \href{https://arxiv.org/abs/1610.04802}{{\ttfamily arXiv:1610.04802}}.

\bibitem[Esteban et~al.(2019)Esteban, Gonzalez-Garcia, Hernandez-Cabezudo,
  Maltoni, and Schwetz]{Esteban:2018azc}
I.~Esteban, M.~Gonzalez-Garcia, A.~Hernandez-Cabezudo, M.~Maltoni, and
  T.~Schwetz, ``{Global analysis of three-flavour neutrino oscillations:
  synergies and tensions in the determination of $\theta_{23}$, $\delta_{CP}$,
  and the mass ordering}'', {\em JHEP} {\bfseries 01} (2019) 106,
  \href{https://arxiv.org/abs/1811.05487}{{\ttfamily arXiv:1811.05487}}.

\bibitem[Huber et~al.(2005)Huber, Lindner, and Winter]{Huber:2004ka}
P.~Huber, M.~Lindner, and W.~Winter, ``{Simulation of long-baseline neutrino
  oscillation experiments with GLoBES (General Long Baseline Experiment
  Simulator)}'', {\em Comput. Phys. Commun.} {\bfseries 167} (2005) 195,
 \href{https://arxiv.org/abs/hep-ph/0407333}{{\ttfamily arXiv:hep-ph/0407333}}.

\bibitem[Huber et~al.(2007)Huber, Kopp, Lindner, Rolinec, and
  Winter]{Huber:2007ji}
P.~Huber, J.~Kopp, M.~Lindner, M.~Rolinec, and W.~Winter, ``{New features in
  the simulation of neutrino oscillation experiments with GLoBES 3.0: General
  Long Baseline Experiment Simulator}'', {\em Comput. Phys. Commun.} {\bfseries
  177} (2007) 432--438,
 \href{https://arxiv.org/abs/hep-ph/0701187}{{\ttfamily arXiv:hep-ph/0701187}}.

\bibitem[Huber and Schwetz(2004)]{Huber:2004xh}
P.~Huber and T.~Schwetz, ``{Precision spectroscopy with reactor
  anti-neutrinos}'', {\em Phys. Rev.} {\bfseries D70} (2004) 053011,
 \href{https://arxiv.org/abs/hep-ph/0407026}{{\ttfamily arXiv:hep-ph/0407026}}.

\bibitem[Vogel and Beacom(1999)]{Vogel:1999zy}
P.~Vogel and J.~F. Beacom, ``The angular distribution of the neutron inverse
  beta decay, $\overline{\nu}_e + p \to e^+ + n$'', {\em Phys. Rev.} {\bfseries
  D60} (1999) 053003,
 \href{https://arxiv.org/abs/hep-ph/9903554}{{\ttfamily hep-ph/9903554}}.

\bibitem[An et~al.(2015)]{An:2015rpe}
{\bfseries Daya Bay} Collaboration, F.~P. An {\em et~al.}, ``{New Measurement
  of Antineutrino Oscillation with the Full Detector Configuration at Daya
  Bay}'', {\em Phys. Rev. Lett.} {\bfseries 115} (2015), no.~11, 111802,
 \href{https://arxiv.org/abs/1505.03456}{{\ttfamily arXiv:1505.03456}}.

\bibitem[An et~al.(2016)]{An:2015jdp}
{\bfseries JUNO} Collaboration, F.~An {\em et~al.}, ``{Neutrino Physics with
  JUNO}'', {\em J.\ Phys.\ G} {\bfseries 43} (2016), no.~3, 030401,
  \href{https://arxiv.org/abs/1507.05613}{{\ttfamily arXiv:1507.05613}}.

\bibitem[Aartsen et~al.(2020)]{Bezerra:2019dao}
{\bfseries IceCube-Gen2, JUNO} Collaboration, M.~Aartsen {\em et~al.},
  ``{Combined sensitivity to the neutrino mass ordering with JUNO, the IceCube
  Upgrade, and PINGU}'', {\em Phys.\ Rev.\ D} {\bfseries 101} (2020), no.~3,
  032006,  \href{https://arxiv.org/abs/1911.06745}{{\ttfamily
  arXiv:1911.06745}}.

\bibitem[Li et~al.(2016)Li, Wang, and Xing]{Li:2016txk}
Y.-F. Li, Y.~Wang, and Z.-z. Xing, ``{Terrestrial matter effects on reactor
  antineutrino oscillations at JUNO or RENO-50: how small is small?}'', {\em
  Chin.\ Phys.\ C} {\bfseries 40} (2016), no.~9, 091001,
  \href{https://arxiv.org/abs/1605.00900}{{\ttfamily arXiv:1605.00900}}.

\bibitem[Khan et~al.(2020)Khan, Nunokawa, and Parke]{Khan:2019doq}
A.~N. Khan, H.~Nunokawa, and S.~J. Parke, ``{Why matter effects matter for
  JUNO}'', {\em Phys.\ Lett.\ B} {\bfseries 803} (2020) 135354,
  \href{https://arxiv.org/abs/1910.12900}{{\ttfamily arXiv:1910.12900}}.

\bibitem[Abrah\~ao et~al.(2015)Abrah\~ao, Minakata, Nunokawa, and
  Quiroga]{Abrahao:2015rba}
T.~Abrah\~ao, H.~Minakata, H.~Nunokawa, and A.~A. Quiroga, ``{Constraint on
  Neutrino Decay with Medium-Baseline Reactor Neutrino Oscillation
  Experiments}'', {\em JHEP} {\bfseries 11} (2015) 001,
  \href{https://arxiv.org/abs/1506.02314}{{\ttfamily arXiv:1506.02314}}.

\bibitem[Porto-Silva et~al.(2020)Porto-Silva, Prakash, Peres, Nunokawa, and
  Minakata]{Porto-Silva:2020gma}
Y.~P. Porto-Silva, S.~Prakash, O.~Peres, H.~Nunokawa, and H.~Minakata,
  ``{Constraining visible neutrino decay at KamLAND and JUNO}'',
  \href{https://arxiv.org/abs/2002.12134}{{\ttfamily arXiv:2002.12134}}.

\bibitem[Khan et~al.(2013)Khan, McKay, and Tahir]{Khan:2013hva}
A.~N. Khan, D.~W. McKay, and F.~Tahir, ``{Sensitivity of medium-baseline
  reactor neutrino mass-hierarchy experiments to nonstandard interactions}'',
  {\em Phys. Rev.} {\bfseries D88} (2013) 113006,
 \href{https://arxiv.org/abs/1305.4350}{{\ttfamily arXiv:1305.4350}}.

\bibitem[Ohlsson et~al.(2014)Ohlsson, Zhang, and Zhou]{Ohlsson:2013nna}
T.~Ohlsson, H.~Zhang, and S.~Zhou, ``{Nonstandard interaction effects on
  neutrino parameters at medium-baseline reactor antineutrino experiments}'',
  {\em Phys. Lett.} {\bfseries B728} (2014) 148--155,
 \href{https://arxiv.org/abs/1310.5917}{{\ttfamily arXiv:1310.5917}}.

\bibitem[Bakhti and Farzan(2014)]{Bakhti:2014pva}
P.~Bakhti and Y.~Farzan, ``{Shedding light on LMA-Dark solar neutrino solution
  by medium baseline reactor experiments: JUNO and RENO-50}'', {\em JHEP}
  {\bfseries 07} (2014) 064,
 \href{https://arxiv.org/abs/1403.0744}{{\ttfamily arXiv:1403.0744}}.

\bibitem[Li and Zhao(2014)]{Li:2014rya}
Y.-F. Li and Z.-h. Zhao, ``{Tests of Lorentz and CPT Violation in the Medium
  Baseline Reactor Antineutrino Experiment}'', {\em Phys. Rev.} {\bfseries D90}
  (2014), no.~11, 113014,
 \href{https://arxiv.org/abs/1409.6970}{{\ttfamily arXiv:1409.6970}}.

\bibitem[Liao et~al.(2017)Liao, Marfatia, and Whisnant]{Liao:2017awz}
J.~Liao, D.~Marfatia, and K.~Whisnant, ``{Nonstandard interactions in solar
  neutrino oscillations with Hyper-Kamiokande and JUNO}'', {\em Phys. Lett.}
  {\bfseries B771} (2017) 247--253,
 \href{https://arxiv.org/abs/1704.04711}{{\ttfamily arXiv:1704.04711}}.

\bibitem[Anamiati et~al.(2019)Anamiati, De~Romeri, Hirsch, Ternes, and
  T\'ortola]{Anamiati:2019maf}
G.~Anamiati, V.~De~Romeri, M.~Hirsch, C.~A. Ternes, and M.~T\'ortola,
  ``{Quasi-Dirac neutrino oscillations at DUNE and JUNO}'', {\em Phys. Rev.}
  {\bfseries D100} (2019), no.~3, 035032,
 \href{https://arxiv.org/abs/1907.00980}{{\ttfamily arXiv:1907.00980}}.

\end{thebibliography}

\begingroup\raggedright\endgroup

\end{document}